\documentclass[12pt]{article}

\usepackage{amsfonts,amsmath,amssymb,verbatim}

\newcommand{\be}{\begin{equation}}
\newcommand{\ee}{\end{equation}}
\newcommand{\ba}{\begin{eqnarray}}
\newcommand{\ea}{\end{eqnarray}}

\begin{document}

\vspace*{1cm}
\begin{center}
{\bf Finite-Size Corrections of the $\mathbb{CP}^3$ Giant Magnons: the L\"{u}scher terms}\\
\vspace{1.8cm} {\large Diego Bombardelli and Davide Fioravanti
\footnote{E-mail:bombardelli@bo.infn.it, fioravanti@bo.infn.it}}\\
\vspace{.5cm} {\em Sezione INFN di Bologna, Dipartimento di Fisica, Universit\`a di Bologna, \\
Via Irnerio 46, Bologna, Italy} \\
\end{center}

\begin{abstract}
{\noindent We compute classical and first quantum finite-size corrections to the recently found giant magnon solutions in two different subspaces of $\mathbb{CP}^3$. We use the L\"{u}scher approach on the recently proposed exact $S$-matrix for $\mathcal{N}=6$ superconformal Chern-Simons theory. We compare our results with the string and algebraic curve computations and find agreement, thus providing a non-trivial test for the new $AdS_4/CFT_3$ correspondence within an integrability framework.}
\end{abstract}

\newpage

\section{Introduction}
In recent years, a great development in the $AdS_5/CFT_4$ correspondence \cite{MGKPW} was put forward thanks to the discovery of integrable structures in both sides of this gauge/string duality (cf. for instance the seminal papers \cite{L, MZ0, BPR}).

Very recently, a new conjecture has been proposed regarding a correspondence between a large $N$ M-theory on $AdS_4\times S^7/\mathbb{Z}_k$ and a three-dimensional $SU(N)\times SU(N)$ Chern-Simons matter theory whith $\mathcal{N}=6$ superconformal symmetry \cite{ABJM}.

Moreover, Minahan and Zarembo \cite{MZ} have shown that this theory is integrable (see also \cite{BR}) at the second order in $\lambda$, that is a 't Hooft coupling defined as $\lambda=N/k$, which is made continuous when $N,k\rightarrow\infty$ and $\lambda$ is kept fixed. On the string theory side, the integrability at the classical level has been shown in \cite{AF0, S}. 

Furthermore, giant magnon \cite{HM} solutions were found in the IIA string theory in $AdS_4\times \mathbb{CP}^3$, dual to the $SU(2)\times SU(2)$ sector of the gauge theory \cite{NT, GGY, GHO, BT}, and their finite-size effects were studied in \cite{GHOS, APGHO, AB, IS, ABR, LPP}. 

On the other hand, an all-loop generalisation of the two-loop Bethe Ansatz proposed in \cite{MZ} was conjectured by \cite{GV} and  may be derived starting from a $S$-matrix proposed in \cite{AN} \footnote{See also \cite{MM} for the original derivation of the SYM Bethe Ansatz equations from the $AdS_5/CFT_4$ $S$-matrix}. Actually, the string theory computations \cite{MR, AAB, K} of the folded spinning string energy led to a result that was different from the Bethe Ansatz prediction of \cite{GV}. It was suggested in \cite{GM} that this disagreement was due to different regularisations used in the calculation of the one-loop correction to the string energy by the algebraic curve method. A resolution to this apparent contradiction between the worldsheet and the Bethe Ansatz calculations has been recently proposed in \cite{MRT}, indicating that the Bethe Ansatz proposal may be correct at strong coupling. Since the dispute concerns one-loop results, we expect agreement in derivations at leading order at strong coupling, in which we are mainly interested in this paper.

In fact, the aim of this paper is to compute leading finite-size corrections in the simplest case of an elementary giant magnon (GM) through the generalised L\"{u}scher method (see \cite{Luscher, AJK, JL, GSV, HJL, GSV1} regarding mainly $AdS_5/CFT_4$), based on scattering data ($S$-matrix).

These terms correct the infinite volume dispersion relation and should take into account possible wrapping effects (cf., for instance, \cite{AJK, RSS} for the more studied phenomenon of $AdS_5/CFT_4$ wrapping).

After the careful analysis in \cite{GGY, GHO}, one can easily understand that there is a classical solution of GM kind in $\mathbb{CP}^3$, which lives in $\mathbb{R}_t\times S^2\times S^2$ and whose infinite volume dispersion relation behaves at large $\lambda$ in this way
\be
\epsilon(p)\simeq2\sqrt{2\lambda}\left|\sin\left(\frac{p}{2}\right)\right|\ .
\label{classical}
\ee
This solution was interpreted in \cite{GHO, GHOS} as composed by two magnons, each one in a $S^2$, with equal worldsheet momenta $p\equiv p_1=p_2$ and the following infinite volume dispersion relation \footnote{We ought to thank G. Grignani for clarifying this point to us.}
\be
\epsilon_{s}(p)=\sqrt{\frac{1}{4}+4h^2(\lambda)\sin^2\left(\frac{p}{2}\right)}\ ,
\label{dispersion}
\ee
where
\be
h(\lambda)=\left\{\begin{array}{l}
\lambda+O(\lambda^2)\ \mbox{for}\ \lambda\ll1\\\\
\sqrt{\lambda/2}+O(\lambda^0)\ \mbox{for}\ \lambda\gg1\ .
\end{array}\right.
\label{h}
\ee
Consistently, at large $\lambda$, (\ref{dispersion}) becomes one half of (\ref{classical}).

Therefore, in our calculations of the L\"{u}scher terms, we will have to use the formulae for multiparticle states (see \cite{OM, VP, HS, BJ} for some applications in $AdS_5/CFT_4$).

On the other hand, \cite{IS} found, by algebraic curve methods, the first quantum correction to the energy of a GM that lives on $\mathbb{CP}^1\approx S^2$, with the same dispersion relation of (\ref{dispersion}). In order to distinguish this solution, we will call it "small" GM.

We will show the calculations for the $\mu$- and $F$-term of the  GM in Section 2 and 3, respectively. We find agreement with the string results for the $\mu$-term and propose a new result for the first quantum finite-size correction, that is very similar, at the level of the final integral expression, to the algebraic curve result for the "big" GM in \cite{IS}. In Section 4, we will present computations for the $\mu$- and $F$-term of the "small" GM. In the latter case we will give a result which confirms the algebraic curve calculations \cite{IS}, for the classical leading contribution, instead, we propose a new result that probably will require a deeper understanding.
Finally, we also give some results for the next-to-leading contributions to the $\mu$-terms in Section 5. We conclude with some conclusions in Section 6.

\section{The $\mu$-term for the $\mathbb{R}\times S^2\times S^2$ giant magnon}

In this section we want to compute the leading finite-size correction to $\epsilon(p)$, $\delta\epsilon^{\mu}(p)$, as the L\"{u}scher $\mu$-term for a nonrelativistic theory characterised by a dispersion relation of kind (\ref{dispersion}).
The generalisation of the L\"{u}scher $\mu$-term energy correction \cite{Luscher} for a single particle to a generic nonrelativistic theory, has been first derived by \cite{JL}, and reads
\be
\delta\epsilon_a^{\mu}=-i\left(1-\frac{\epsilon'(p)}{\epsilon'(\tilde{q}^*)}\right)e^{-i\tilde{q}^*L}\,\mathop{Res}\limits_{q^*=\tilde{q}^*}\sum_b(-1)^{F_b}S_{ba}^{ba}(q^*,p)\ ,
\label{mu-term}
\ee
where $\tilde{q}^*$ corresponds to the bound state pole of the $S$-matrix, $p$ is the momentum of the real  particle, denoted by $a$.
The on-shell condition for the virtual particle imposes 
\be
q^2+\epsilon^2(q^*)=0\ ,
\ee
which entails the inverse relation 
\be
q^*=-2\,i\,\mbox{arcsinh}\frac{\sqrt{1+4\,q^2}}{4\,h(\lambda)}\ .
\ee
If we define, as, for instance, in \cite{GV}, the variables $x^{\pm}$ such that
\be
\frac{x^+}{x^-}=e^{ip/2}\,;\ \ \ x^{+}+\frac{1}{x^{+}}-x^{-}-\frac{1}{x^{-}}=\frac{i}{h(\lambda)}\ ,
\label{constraint}
\ee
then we have explicit expressions of $x^{\pm}$ in terms of the momentum:
\be
x_p^{\pm}=e^{\pm ip/2}\,\frac{1+\sqrt{1+16\,h^2(\lambda)\sin^2\left(\frac{p}{2}\right)}}{4\,h(\lambda)\sin(p/2)}
\ee
and we can parametrize the energy of one magnon in these two ways
\be
\epsilon_s(p)=i\,h(\lambda)(x^--x^+)-\frac{1}{2}=i\,h(\lambda)\left(\frac{1}{x^+}-\frac{1}{x^-}\right)+\frac{1}{2}\ .
\ee
Therefore, the asymptotic expansions at strong coupling for $x^{\pm}$ are very similar to those in \cite{JL}:
\ba
x_p^+&=&e^{ip/2}\left(1+\frac{1}{2\sqrt{2\lambda}\sin(p/2)}+O\left(\frac{1}{\lambda}\right)\right)\nonumber\\
x_p^-&=&e^{-ip/2}\left(1+\frac{1}{2\sqrt{2\lambda}\sin(p/2)}+O\left(\frac{1}{\lambda}\right)\right)\ .
\label{asymp}
\ea
Now, since the GM solution on $\mathbb{R}_t\times S^2\times S^2$ was interpreted in \cite{GHO, GHOS} as a couple of two magnons with equal momenta, then, in order to calculate the finite-size correction, we have to reconsider generalised L\"{u}scher formulae for multiparticle states \cite{HS} rather than (\ref{mu-term}):
\be
\delta\epsilon_A^{\mu}=-i\sum_{l=1}^M\sum_b(-1)^{F_b}\left(1-\frac{\epsilon'_{a_l}(p_l)}{\epsilon'_{b}(\tilde{q}^*_l)}\right)e^{-i\tilde{q}^*_lL}\,\mathop{Res}\limits_{q^*=\tilde{q}^*_l}S_{ba_l}^{ba_l}(q^*,p_l)\prod_{k\neq l}^MS_{ba_k}^{ba_k}(\tilde{q}^*_l,p_k)\ ,
\label{multimu-term}
\ee
where $A\equiv\left\lbrace a_1(p_1),...,a_M(p_M)\right\rbrace$ denotes a string made of $M$ GMs.

We have to apply this formula to the case of two real particles - one of type A and the other of type B - interacting with another couple of virtual particles - of type A and B - moving around the cylinder.
The $S$-matrices we will use to describe these interactions - between A-A, A-B and B-B particles - are those proposed in \cite{AN}:
\ba
S^{AA}(p_1,p_2)&=&S^{BB}(p_1,p_2)=S_0(p_1,p_2)\hat{S}(p_1,p_2)=\sigma(p_1,p_2)\,\frac{1-\frac{1}{x_1^+x_2^-}}{1-\frac{1}{x_1^-x_2^+}}\,\hat{S}(p_1,p_2)\ ;\nonumber\\ S^{AB}(p_1,p_2)&=&S^{BA}(p_1,p_2)=\tilde{S}_0(p_1,p_2)\hat{S}(p_1,p_2)=\sigma(p_1,p_2)\,\frac{x_1^--x_2^+}{x_1^+-x_2^-}\,\hat{S}(p_1,p_2)\ ,
\label{s-matrices}
\ea
where $\sigma(p_1,p_2)$ is the BES/BHL dressing factor \cite{AFS, BES, HL, BHL}, and $\hat{S}$ can expressed just as in the appendix A.5 of \cite{AFZ}, through a set of functions $a_1,...,a_{10}$ dependent on the variables $x_{1,2}^{\pm}$.
Moreover, we take into account that only $S^{AA}$, $S^{BB}$ have physical poles, corresponding to BPS bound states, determined by the same condition $x_q^-=x_p^+$. Then the l.h.s. of this equation espands as
\be
x_q^-=e^{ip/2}\left(1+\frac{1}{2\sqrt{2\lambda}\sin(p/2)}+O\left(\frac{1}{\lambda}\right)\right)
\ee
and the corresponding $x_q^+$ can be found from inserting this into the second equation of (\ref{constraint}), and is thus given by
\be
x_q^+=e^{ip/2}\left(1+\frac{3}{2\sqrt{2\lambda}\sin(p/2)}+O\left(\frac{1}{\lambda}\right)\right)\ .
\ee
In other words, the real A-particle scatters with the virtual A-particle and forms with it a bound state - corresponding to the physical pole of $S^{AA}$ - while the B-particle scatters elastically with the virtual A-particle, because $S^{AB}$ does not have physical poles, as we will see below. Then the real B-particle proceeds to form a bound state - corresponding to the physical pole of $S^{BB}$ - with the virtual B-particle. Finally, we have to sum the contributions given by this diagram over all the possible residues of the $S$-matrix, taking into account that both real paricles belong to the SU(2) sector (i.e. $a_1=a_2=1$).
Moreover, in this case the real particles have equal momenta $p_1=p_2=p$, then $S^{AA}(q^*,p)$ and $S^{BB}(q^*,p)$ share the same pole. Hence, the above description shall suggest us this $S$-matrix contribution:

\be
S^{AA}(q^*,p)S^{AB}(q^*,p)+S^{BA}(q^*,p)S^{BB}(q^*,p)\ ,
\ee
on which we need to pick up the residues. Then we may propose the following expression for the $\mu$-term of the SU(2)$_A\times$ SU(2)$_B$ giant magnon

\ba
\label{2mu-term}
\delta\epsilon^{\mu}&=&-i\sum_b(-1)^{F_b}\left\lbrace\left(1-\frac{\epsilon'_{1}(p)}{\epsilon'_{b}(\tilde{q}^*_1)}\right)e^{-i\tilde{q}^*_1L}\left[(S^{AB})_{b1}^{b1}(\tilde{q}_1^*,p)\mathop{Res}\limits_{q^*=\tilde{q}^*_1}(S^{AA})_{b1}^{b1}(q^*,p)+\right.\right.\nonumber\\
&+&\left.(S^{BA})_{b1}^{b1}(\tilde{q}^*_1,p)\mathop{Res}\limits_{q^*=\tilde{q}^*_1}(S^{BB})_{b1}^{b1}(q^*,p)\right]+\left(1-\frac{\epsilon'_{1}(p)}{\epsilon'_{b}(\tilde{q}^*_2)}\right)e^{-i\tilde{q}^*_2L}\times\\
&\times&\left.\left[(S^{AB})_{b1}^{b1}(\tilde{q}^*_2,p)\mathop{Res}\limits_{q^*=\tilde{q}^*_2}(S^{AA})_{b1}^{b1}(q^*,p)+(S^{BA})_{b1}^{b1}(\tilde{q}^*_2,p)\mathop{Res}\limits_{q^*=\tilde{q}^*_2}(S^{BB})_{b1}^{b1}(q^*,p)\right]\right\rbrace\nonumber\ .
\ea
Furthermore, the asymptotics of the pole momentum $\tilde{q}^*_{1}=\tilde{q}^*_{2}\equiv\tilde{q}^*$ is the same and rather similar to that of the SYM case \cite{JL}, and gives the expression for the exponential term in (\ref{2mu-term}):
\be
e^{-i\tilde{q}^*L}\simeq e^{-\frac{L}{\sqrt{2\lambda}\sin(p/2)}}\ .
\ee
Also, the two kinematical factors in (\ref{2mu-term}) give exactly the same contibution, whose leading order at strong coupling is obviously identical to that in the SYM case \footnote{It depends on the ratio of first derivatives of the dispersion relation, that can be seen as one half of the SYM giant magnon relation, after substituting $g$ for $h(\lambda)$, where $g=\sqrt{\lambda_{SYM}}/4\pi$.}:
\be
1-\frac{\epsilon'_1(p)}{\epsilon'_{b}(\tilde{q}^*)}=\sin^2\left(\frac{p}{2}\right)+O\left(\frac{1}{\sqrt{\lambda}}\right)\ .
\label{kin}
\ee
Therefore, the expression (\ref{2mu-term}) becomes
\be
\delta\epsilon^{\mu}\simeq-4\,i\sin^2\left(\frac{p}{2}\right)e^{-\frac{L}{\sqrt{2\lambda}\sin(p/2)}}\,\sum_b(-1)^{F_b}\left[(S^{AB})_{ba}^{ba}(\tilde{q}^*,p)\mathop{Res}\limits_{q^*=\tilde{q}^*}(S^{AA})_{ba}^{ba}(q^*,p)\right]\ ,
\label{mutot}
\ee
where the factor 4 comes out from the fact that $S^{AA}$ ($S^{AB}$) is equal to $S^{BB}$ ($S^{BA}$), as written in (\ref{s-matrices}), and $p_1=p_2=p$.

Now, it remains to evaluate the $S$-matrix contribution. We have taken $a_{1,2}$ in the SU(2) sector and $b=1,...,4$, then we select only certain elements on the diagonal of the SU(2$|$2) $S$-matrix.
Therefore, as for the SU(2) sector of SYM, the $S$-matrix contribution reads
\ba
\label{resAA}
&&\mathop{Res}\limits_{q^*=\tilde{q}^*}\left(S^{AA}\right)_{b1}^{b1}(q^*,p)=\mathop{\lim}\limits_{q^*\rightarrow\tilde{q}^*}\left(\frac{q^*-\tilde{q}^*}{x_q^--x_p^+}\right)\frac{1-\frac{1}{x_q^+x_p^-}}{1-\frac{1}{x_q^-x_p^+}}\,(x_q^--x_p^+)\Big\lbrace a_1(x_q,x_p) E^1_1 \otimes E^1_1\nonumber\\
&&+\left[a_1(x_q,x_p) + a_2(x_q,x_p)\right] E^1_1\otimes E^2_2 + a_6(x_q,x_p) (E^1_1 \otimes E^3_3 + E^1_1 \otimes E^4_4) \Big\rbrace\sigma(x_q,x_p)\nonumber\\
&&\left(S^{AB}\right)_{b1}^{b1}(\tilde{q}^*,p)=\frac{x_{q}^--x_{p}^+}{x_{q}^+-x_{p}^-}\Big\lbrace a_1(x_{q},x_{p}) E^1_1 \otimes E^1_1+\left[a_1(x_{q},x_{p})+ a_2(x_{q},x_{p})\right] E^1_1\otimes E^2_2\nonumber\\
&&+a_6(x_{q},x_{p}) (E^1_1 \otimes E^3_3 + E^1_1 \otimes E^4_4) \Big\rbrace\sigma(x_{q},x_{p})\ .
\ea
Following the calculations of \cite{JL} for the strong coupling limit, one obtains, for the limit in the first line of (\ref{resAA})
\be
\mathop{\lim}\limits_{q^*\rightarrow\tilde{q}^*}\left(\frac{q^*-\tilde{q}^*}{x_q^--x_p^+}\right)=\frac{1}{x_q^{-'}}=\frac{i\,e^{-i\frac{p}{2}}}{2\sin^2\left(\frac{p}{2}\right)}+O\left(\frac{1}{\sqrt{\lambda}}\right)\ .
\label{1/xq'}
\ee
On the other hand, for the remaining "undressed" part we have that only the term involving $a_1(x_q,x_p)$ survives at strong coupling:
\be
2\,\frac{1-\frac{1}{x_q^+x_p^-}}{1-\frac{1}{x_q^-x_p^+}}\,\frac{(x_{q}^--x_{p}^+)^2}{x_{q}^+-x_{p}^-}\,a_1^2(x_q,x_p)= \frac{2\sqrt{2}\,e^{3i\frac{p}{2}}}{\sqrt{\lambda}\sin\left(\frac{p}{2}\right)}+O\left(\frac{1}{\lambda}\right)\ ,
\ee
and then
\be
\sum_b(-1)^{F_b}(S^{AB})_{ba}^{ba}(\tilde{q}^*,p)\mathop{Res}\limits_{q^*=\tilde{q}^*}(S^{AA})_{ba}^{ba}(q_*,p)\simeq\frac{i\sqrt{2}\,e^{ip}}{\sqrt{\lambda}\sin^3\left(\frac{p}{2}\right)}\,\sigma(x_{q},x_p)^2\ .
\ee  
The contribution of the dressing factor is given by
\be
\sigma(x_q,x_p)=e^{i\left[\chi(x_p^+,x_q^+)+\chi(x_p^-,x_q^-)-\chi(x_q^+,x_p^-)\right]}= \frac{\sqrt{2\lambda}}{e}\,i\,e^{-i\frac{p}{2}}\sin^2\left(\frac{p}{2}\right)+O\left(\lambda^0\right)\ ,
\label{dressing}
\ee
where $\chi(x,y)$ is defined as in \cite{DHM} 
\be
\chi(x,y)=\chi_{AFS}(x,y)+\chi_{HL}(x,y)+\sum_{n=2}^{\infty}\chi^{(n)}(x,y)
\ee
and is completely identical to the definitions in \cite{JL}, except for
\be
\chi_{AFS}(x,y)=-h(\lambda)\left(\frac{1}{y}-\frac{1}{x}\right)\left[1-(1-xy)\ln\left(\frac{1}{xy}\right)\right]\ ,
\ee
where we have mapped $g_{Ref. \cite{JL}}/\sqrt{2}$ to $h(\lambda)$.

Finally, the dressing factor total contribution is given by (similarly to \cite{JL}) 
\be
\sigma(x_{q},x_p)^2=-\frac{2\lambda}{e^2}\sin^4\left(\frac{p}{2}\right)e^{-ip}+O\left(\sqrt{\lambda}\right)\ ,
\ee
so that the $\mu$-term (\ref{mutot}) becomes
\ba
\delta\epsilon^{\mu}&\simeq&-4\,i\sin^2\left(\frac{p}{2}\right)e^{-\frac{L}{\sqrt{2\lambda}\sin(p/2)}}\,\frac{i\sqrt{2}\,e^{ip}}{\sqrt{\lambda}\sin^3\left(\frac{p}{2}\right)}\,\sigma(x_{q},x_p)^2\ .
\ea
In conclusion, reassembling all these contributions, we obtain the result for the finite-size correction to the dispersion relation of a GM in $SU(2)\times SU(2)$, in perfect agreement with that given in equations (2) and (54) of \cite{GHOS}:
\be
\delta\epsilon^{\mu}\simeq-\frac{8\sqrt{2\lambda}}{e^2}\sin^3\left(\frac{p}{2}\right)e^{-\frac{L}{\sqrt{2\lambda}\sin(p/2)}}\ .
\label{solmu}
\ee
Now, we shall compare this to that obtained for the SYM case. There, it has been obtained \cite{AFZ1, AFGS, JL, AF}
\be
\delta\epsilon^{\mu}_{SYM}\simeq-\frac{16}{e^2}\,g\sin^3\left(\frac{p}{2}\right)e^{-\frac{L}{\sqrt{2}g\sin(p/2)}}\ .
\ee
Hence, it can be directly related to (\ref{solmu}) simply substituting $g$ with $h(\lambda)\simeq\sqrt{\lambda/2}$. This is exactly the map that relates the dispersion relation (\ref{classical}) to the analogue in SYM.

\section{The $F$-term for the $\mathbb{R}\times S^2\times S^2$ giant magnon}

The formula for the one-particle L\"{u}scher $F$-term \cite{JL} is
\be
\delta E_{a}^F= - \int_{-\infty}^{+\infty} \frac{dq}{2\pi}\left(1- \frac{\epsilon' (p)}{\epsilon' (q^*)} \right) \, e^{-i q^* L} \sum_b (-1)^{F_b}\left( S_{b a}^{\, b a} (q^*, p)-1 \right)\ .
\ee
Here we consider - as in the previous Section - a GM with excitations on both $S^2$; then we have to take into account interactions between A and B particles inserting also $S^{AB}$ in the final $S$-matrix.
The generalised multi-particle formula for the L\"{u}scher $F$-term is \cite{HS}
\be
\delta E_A^F= - \sum_b (-1)^{F_b} \int_{-\infty}^{+\infty} \frac{dq}{2\pi}\left(1- \sum_{k=1}^M \alpha_k \, \frac{\epsilon_{a_k}' (p_k)}{\epsilon_b' (q^*)} \right) \, e^{-i q^* L} \left( \prod_{l=1}^M S_{b a_l}^{\, b a_l} (q^*, p_l)-1 \right)\ ,
\label{multi F-term}
\ee
where $\sum_{k=1}^M \alpha_k=1$.

We start from determining the kinematic part of the integral above. Firstly, the energy of the virtual particle with momentum $q^*$ is parametrised by the variables $x_{q^*}^{\pm}$, that scale at strong coupling as \cite{GV}
\be
x_{q^*}^{\pm}=x\pm\frac{i\,x^2}{2\,h(\lambda)(x^2-1)}+O(1/\lambda)\ .
\label{xpm}
\ee
In SYM, following the notation of \cite{GSV}, this corresponds to
\be
X_q^{\pm}=x\pm\frac{4\,\pi\,x^2}{\sqrt{\lambda}\,(x^2-1)}+O(1/\lambda)\ .
\ee
Now, we can write the infinite volume dispersion relation for the virtual particle and its expression at strong coupling in terms of $x$
\be
\epsilon(q^*)=\frac{h(\lambda)}{2\,i}\left(x_{q^*}^+-\frac{1}{x_{q^*}^+}-x_{q^*}^-+\frac{1}{x_{q^*}^-}\right)\simeq\frac{x^2+1}{2\,(x^2-1)}\ .
\ee
Moreover, we have to impose the on-shell condition $q^2+\epsilon^2(q^*)=0$, so that we have the following expression for $q$ and $q_*$ in terms of $x$:
\be
q\simeq i\,\frac{x^2+1}{2\,(x^2-1)}\,;\ \ \ q^*\simeq\frac{\sqrt{2}\,x}{\sqrt{\lambda}\,(x^2-1)}\ .
\label{qq*}
\ee
In order to calculate the whole kinematic factor in (\ref{multi F-term}), we first derivative the energy formula in $p$ and in $q^*$ respectively:
\be
\epsilon'(p)=\frac{h(\lambda)}{i}\left(\frac{x_p^++x_p^-}{x_p^+\,x_p^-+1}\right)\,;\ \ \  \epsilon'(q^*)=\frac{h(\lambda)}{i}\left(\frac{2\,x}{x^2+1}\right)\ ,
\ee
then we can now determine
\be
\partial_x\Omega(x)\equiv i\,\frac{dq}{dx}\left(1-\frac{\epsilon'(p)}{\epsilon'(q^*(q))}\right)=\frac{1}{(x^2-1)^2}\left(-2x+(x^2+1)\,\frac{x_p^++x_p^-}{x_p^+\,x_p^-+1}\right)\ ,
\label{cond}
\ee
where $\Omega(x)$ is usually defined in the algebraic curve approach as the function determining the characteristic frequencies of the energy fluctuations. In particular, definition (\ref{cond}) is satisfied by
\be
\Omega(x)=\frac{1}{x^2-1}\left(1-\frac{x_p^++x_p^-}{x_p^+\,x_p^-+1}\,x\right)\ ,
\label{omega}
\ee
which coincides with the expression \cite{IS} valid for both "small" and "big" GM solutions.

As we have proposed in the previous section, the $S$-matrix contribution is given by:
\ba
&&\sum_b(-1)^{F_b}(S^{AA})_{b 1}^{\, b 1} (q^*, p)(S^{AB})_{b 1}^{\, b 1} (q^*, p)+(S^{BB})_{b 1}^{\, b 1} (q^*, p)(S^{BA})_{b 1}^{\, b 1} (q^*, p)=\nonumber\\
&&=2\,S_0(q^*,p)\tilde{S}_0(q^*,p)\left\lbrace a_1^2(x_{q^*},x_p)+\left[a_1(x_{q^*},x_p)+a_2(x_{q^*},x_p)\right]^2-2\,a_6^2(x_{q^*},x_p)\right\rbrace \ .
\label{s-matrix2m}
\ea
In particular, once we insert the above expression into (\ref{multi F-term}) and take the residues in $\tilde{q}^*_1$ and $\tilde{q}^*_2$, we may derive the expression (\ref{2mu-term}) for the $\mu$-term.
The dressing part in (\ref{s-matrix2m}) contributes with
\be
S_0(q^*,p)\tilde{S}_0(q^*,p)=\frac{x_{q^*}^--x^+_{p}}{x_{q^*}^+-x^-_{p}}\,\frac{1-\frac{1}{x_{q^*}^+x^-_p}}{1-\frac{1}{x_{q^*}^-x^+_p}}\,\sigma(x_{q^*},x_p)^2\simeq e^{-\frac{i\sqrt{2}(\Delta-L)}{\sqrt{\lambda}}}\ ,
\ee
where $\Delta=L+2\sqrt{2\lambda}\sin\left(p/2\right)$, while at strong coupling the contribution of $a_2$ can be neglected at leading order, and then the "undressed" part in (\ref{s-matrix2m}) results
\be
2\,\left[a_1(x_{q^*},x_p)^2-a_6(x_{q^*},x_p)^2\right]\simeq 2\left[\left(\frac{x-x^-_p}{xx^-_p-1}\right)^2-1\right]\simeq \frac{4\,i\,e^{-ip/2}\sin(p/2)(x^2-1)}{\left(e^{ip/2}-x\right)^2}\ ,
\ee
where the sign minus of the term $a_6(x_q,x_p)^2$ is due to the term $(-1)^{F_b}$, with $F_b=0$ for bosonic and $F_b=1$ for fermionic terms.
 
All together these terms give 
\be
\delta E^F\simeq 2\oint_{\mathbb{U}^+}\,\frac{dx}{\pi i}\partial_x\Omega(x)\,e^{-i\sqrt{2}\frac{\Delta
}{\sqrt{\lambda}}\frac{x}{x^2-1}}\left[\left(\frac{x-x^-_p}{x\,x^-_p-1}\right)^2-1\right]\ ,
\label{intF}
\ee
that, except for the parametrisation in terms of $p/2$ instead of $p/4$, coincides with the result of \cite{IS} for the "big" GM; then the result obtained with the algeraic curve and with the L\"{u}scher techniques seems to be formally in agreement at all orders in $\sqrt{\lambda}/L$ (the situation here is exactly the same as in the $AdS_5/CFT_4$ correspondence \cite{GSV}).
Using the saddle point method, we can give an approximated evaluation of the above integral at leading order in $\sqrt{\lambda}/\Delta
$:
\be
\delta E^F
\simeq-4\,\sqrt{\frac{\sqrt{\lambda/2}}{\pi\Delta
}}\,e^{-\frac{\Delta
}{\sqrt{2\lambda}}}\,\frac{\sin\left(\frac{p}{2}\right)}{1-\sin\left(\frac{p}{2}\right)}\ .
\label{solF}
\ee
We immediately notice that our result (\ref{solF}), even after a replacement $p/2\rightarrow p/4$, is different to the same quantity (41) in \cite{IS}, although the integral expression for the $F$-term in the line before is the same in that paper \footnote{Except the dependence in $p$, once we replace $E_{Ref. \cite{IS}}(Q=1)$ by our $\Delta$, the two results match at strong coupling.}. We think that the reason of this discrepancy is the different integration curve in the complex plane adopted here. Here we have integrated only on the upper half of the unite circle, because the bijective map $q\rightarrow x$ in (\ref{qq*}) send the real axis to the upper half-circle, as explained in \cite{GSV}. Therefore, we think that there is a mistake in \cite{IS} about the evaluation of this integral. We will explain this point in the Appendix in a more detailed way.

Contrarily to the previous case, the dependence on $p$ is very different if compared with the $F$-term of two GMs in SYM \cite{GSV}, which reads
\be
\delta E^F_{SYM}\simeq-16\,\sqrt{\frac{g}{\pi\Delta_2^{SYM}}}\,e^{-\frac{\Delta_2^{SYM}}{2g}}\left(\frac{\sin^2\left(\frac{p}{2}\right)}{\left(\sin^2\left(\frac{p}{2}\right)-1\right)^2}\right)\ ,
\ee
where $\Delta_2^{SYM}=L+8\,g\sin(p/2)$. 
Only the $p$ independent part, i.e. prefactor and exponential term could reproduce our result (\ref{solF}) by the substitution $g\rightarrow h(\lambda)$ and taking the strong coupling expression. Besides, the expressions of the $\Delta$s would differ for a factor 2 in front of the coupling constant.

\section{The $\mu$- and the $F$-term of the $\mathbb{CP}^1$ giant magnon}

Let us consider a single giant magnon which belongs to the SU(2)$_A$ sector, for instance.
If we take the formula (\ref{mu-term}) for the one-particle case and consider the $S$-matrix contribution \footnote{Actually the contribution of $S^{AB}$ into the self-energy processes of a single A-particle, in a system of A-B particles, is considered also in QCD applications of the L\"uscher terms: see, for instance, \cite{KK}.}
\be
S^{AA}(p_1,p_2)+S^{AB}(p_1,p_2)=\left(\frac{1-\frac{1}{x_1^+x_2^-}}{1-\frac{1}{x_1^-x_2^+}}+\frac{x_1^--x_2^+}{x_1^+-x_2^-}\right)\sigma(p_1,p_2)\,\hat{S}(p_1,p_2)\ ,
\label{s-matrix}
\ee
then we have all the ingredients to compute the $\mu$-term of this "small" GM.

Indeed, we have only to repeat the calculations of the Section 2, without considering contributions by $S^{AB}$, but only by $S^{AA}$, since the latter only has a pole corresponding to a boundstate of two A-particles, that is a necessary condition to have a non-vanishing residue of the $S$-matrix, which determines the $\mu$-term:
\be
\sum_b(-1)^{F_b}\mathop{Res}\limits_{q^*=\tilde{q}^*}(S^{AA})_{b1}^{b1}(q_*,p)\simeq\frac{2}{x_q'^-}
\,\frac{1-\frac{1}{x_q^+x_p^-}}{1-\frac{1}{x_q^-x_p^+}}\,(x_q^--x_p^+)\,a_1(x_q,x_p)\,\sigma(x_{q},x_p)\ .
\ee  
Now, we can take the expressions (\ref{1/xq'}) and (\ref{dressing}), multiply them by
\be
2\,\frac{1-\frac{1}{x_q^+x_p^-}}{1-\frac{1}{x_q^-x_p^+}}\,(x_q^--x_p^+)\,a_1(x_q,x_p)= \frac{2\sqrt{2}\,e^{ip}}{\sqrt{\lambda}\sin\left(\frac{p}{2}\right)}+O\left(\frac{1}{\lambda}\right)
\ee
to obtain the complete contribution of the $S$-matrix:
\be
\sum_b(-1)^{F_b}\mathop{Res}\limits_{q^*=\tilde{q}^*}(S^{AA})_{b1}^{b1}(q_*,p)=-\frac{2}{e\sin(p/2)}+O\left(\frac{1}{\sqrt{\lambda}}\right)\ .
\ee
Finally, inserting the expression for the kinematical factor (\ref{kin}), we obtain the $\mu$-term for the "small" GM at strong coupling
\be
\delta\epsilon_{s}^{\mu}\simeq\frac{2\,i}{e}\sin\left(\frac{p}{2}\right)e^{-\frac{L}{\sqrt{2\lambda}\sin(p/2)}}\ ,
\label{musmall}
\ee
that surprisingly is an imaginary quantity! We have to make a remark on this strange fact. At this point we could think that we should take the real part of this result, as stated in \cite{VP, HS}, because of the replacement $\cos(q^* L)\rightarrow e^{-iq^*L}$ made in the derivation of $\mu$- and $F$-term in \cite{Luscher, JL}.
Alternatively, one could instead try to find a formulation - as in \cite{GSV1} form algebraic curve method - that could guarantee the reality of the whole - not expanded in $(1/\sqrt{\lambda})$ - expression (\ref{mu-term}). We hope, but at this moment we cannot demonstrate, that this is the case, and we reserve this problem to future investigations. 

Now, we want to compute the $F$-term leading contibution for this "small" GM solution.
Thus, in this case we can take the whole $S$-matrix contribution (\ref{s-matrix}), so that we obtain
\be
\sum_b(-1)^{F_b}\left[\left(S^{AA}+S^{AB}\right)\right]_{b1}^{b1}(q^*,p)=(2\,a_1+a_2-2\,a_6)\,\sigma(q^*,p)\left(\frac{1-\frac{1}{x_{q*}^+x_p^-}}{1-\frac{1}{x_{q^*}^-x_p^+}}+\frac{x_{q^*}^--x_p^+}{x_{q^*}^+-x_p^-}\right)\ .
\ee
Taking the expressions in the appendix A of \cite{GSV} for the strong coupling limit of the elements $a_1$, $a_2$, $a_6$ and for $\sigma(x_{q^*},x_p)$, and using the expansion (\ref{xpm}) for $x_{q^*}$ and $x^{\pm}_p\simeq e^{\pm ip/2}$, the the previous equation becomes 
\be
\sum_b(-1)^{F_b}\left[\left(S^{AA}\right)_{b1}^{b1}+\left(S^{AB}\right)_{b1}^{b1}\right](q^*,p)\simeq 4\,e^{-i\sqrt{2}\frac{(\Delta_s-L)}{\sqrt{\lambda}}\frac{x}{x^2-1}}\left(\frac{x-x_p^-}{x_p^-\,x-1}-1\right)\ ,
\ee
where $\Delta_s=L+\sqrt{2\lambda}\sin(p/2)$.
In this way we obtain the following expression
\be
\delta E^F_{s}=2\oint_{\mathbb{U}^+}\frac{dx}{\pi i}\,\partial_x\Omega(x)\,e^{-i\sqrt{2}\frac{\Delta_s}{\sqrt{\lambda}}\frac{x}{x^2-1}}\left(\frac{x-x_p^-}{x_p^-\,x-1}-1\right)\ ,
\ee
that agrees with the expression for the one loop finite-size correction of the "small" GM in \cite{IS}.
If we proceed now to evaluate this integral via the usual saddle-point method, we find
\be 
\delta E^F_{s}\simeq-2\,\sqrt{\frac{\sqrt{\lambda/2}}{\pi\Delta_s}}\,e^{-\frac{\Delta_s}{\sqrt{2\lambda}}}\left(\frac{\cos(p/2)}{1-\sin(p/2)}-1\right)\ .
\label{sol1}
\ee
Obviously, the result of \cite{IS} is different from ours in the same way as in the previous section: there is a discrepancy in evaluating the final integral for the $F$-term (more details in the Appendix).

Comparing with the SYM result \cite{GSV}, we can see that our expression for the first quantum correction to the finite- size effect is very different, because of the different dependence on the momentum:
\be
\delta E^F_{SYM}\simeq-8\,\sqrt{\frac{g}{\pi\Delta^{SYM}}}\,e^{-\frac{\Delta^{SYM}}{2g}}\left(\frac{\cos\left(\frac{p}{2}\right)-1}{\sin\left(\frac{p}{2}\right)-1}\right)\ ,
\label{solsym}
\ee 
where $\Delta^{SYM}=L+4\,g\sin(p/2)$. However, the prefactor and the exponential term, as one can easily notice comparing the two expressions (\ref{sol1}) and (\ref{solsym}), can be mapped, except to the specific form of the $\Delta$s, to our result substituting $g$ with the first order term of $h(\lambda)$ at strong coupling.

\section{Next-to-leading contribution of the $\mu$-terms }

While considering the next-to-leading term, predicted to be a constant $c=-\ln(2)/2\pi$ by \cite{MRT}, in the strong coupling expansion of the central function $h(\lambda)$

\be
h(\lambda)=\sqrt{\lambda/2}+c+O\left(\frac{1}{\sqrt{\lambda}}\right)\ \mbox{for}\ \lambda\gg1\ ,
\label{hsub}
\ee
we may proceed to the expansion of the Zhukovsky variables $x_{p,q}^{\pm}$ up to the order $1/\lambda^{3/2}$:

\ba
x_p^{\pm}&=&e^{\pm ip/2}\left(1+\frac{1}{2\sqrt{2\lambda}\sin(p/2)}+\frac{1}{16\lambda\sin^2(p/2)}-\frac{c}{2\lambda\sin(p/2)}+O\left(\frac{1}{\lambda^{3/2}}\right)\right)\\
x_q^+&=&e^{ip/2}\left(1+\frac{3}{2\sqrt{2\lambda}\sin(p/2)}+i\,\frac{17-e^{ip}-48\,c\sin^2(p/2)}{32\,\lambda \,e^{ip/2}\sin^3(p/2)}+O\left(\frac{1}{\lambda^{3/2}}\right)\right)\ ,
\label{asympsub}
\ea
and $x_q^-$ determined by the boundstate condition $x_q^-=x_p^+$.
The exponential term is now given by

\be
e^{-i\tilde{q}^*_{1,2}L}\equiv e^{-i\tilde{q}^*L}= e^{-\frac{L}{\sqrt{2\lambda}\sin(p/2)}}\left[1-\frac{L}{2\lambda}\left(i\,\frac{\cos(p/2)}{\sin^3(p/2)}-\frac{c}{\sin(p/2)}\right)\right]+O\left(\frac{1}{\lambda^{3/2}}\right)\ ,
\ee
while the kinematical factor reads
\be
1-\frac{\epsilon'_{a_{1,2}}(p_{1,2})}{\epsilon'_{b}(\tilde{q}^*)}=\sin^2\left(\frac{p}{2}\right)-\frac{i\,\cos(p/2)}{\sqrt{2\lambda}}+O\left(\frac{1}{\lambda^{3/2}}\right)\ .
\label{kinsub}
\ee

It remains to evaluate the $S$-matrix contribution, that is given in part by the following limit
\be
\mathop{\lim}\limits_{q^*\rightarrow\tilde{q}^*}\left(\frac{q^*-\tilde{q}^*}{x_q^--x_p^+}\right)=\frac{1}{x_q^{-'}}=\frac{i\,e^{-i\frac{p}{2}}}{2\sin^2\left(\frac{p}{2}\right)}-\frac{3+2\,e^{-ip}}{4\sqrt{2\lambda}\sin^4(p/2)}+O\left(\frac{1}{\lambda}\right)\ ,
\label{1/xq'sub}
\ee
that is due to taking the residue in the momentum of the boundstate, in part by the "undressed" $S$-matrix elements

\be
\frac{1-\frac{1}{x_q^+x_p^-}}{1-\frac{1}{x_q^-x_p^+}}\,\frac{(x_{q}^--x_{p}^+)^2}{x_{q}^+-x_{p}^-}\,[a_1^2+(a_1+a_2)^2-2\,a_6^2]= \frac{2\sqrt{2}\,e^{3i\frac{p}{2}}}{\sqrt{\lambda}\sin\left(\frac{p}{2}\right)}+\frac{i\,e^{ip}-4\,c\sin^2(p/2)}{\lambda\sin^3(p/2)}+O\left(\frac{1}{\lambda^{3/2}}\right)
\ee
and in part by the dressing factor

\ba
\sigma^2(x_q,x_p)&=&-\frac{4\,\lambda}{e^2}\,e^{-ip}\sin^4\left(\frac{p}{2}\right)-\frac{2\sqrt{2\lambda}\,e^{-ip}\sin^2(p/2)}{\pi\,e^2}\nonumber\\
&-&\frac{\sqrt{2\lambda}\,e^{-ip}\sin^2(p/2)[4\,c\sin^2(p/2)+\sin(p/2)-5\,i\cos(p/2)]}{e^2}+O\left(\lambda^0\right)\ .
\label{dressingsub}
\ea
Therefore, all these contributions together give in the end

\ba
\delta\epsilon^{\mu}&=&\left[-8\sqrt{2\lambda}\sin^3\left(\frac{p}{2}\right)-\frac{16\sin\left(\frac{p}{2}\right)}{\pi}+8\,i\sin\left(\frac{p}{2}\right)-16\,c\sin^3\left(\frac{p}{2}\right)\right.\nonumber\\
&+&\left.\frac{4\sqrt{2}L}{\sqrt{\lambda}}\left(i\cos\left(\frac{p}{2}\right)-2\,c\sin^2\left(\frac{p}{2}\right)\right)\right]e^{-2-\frac{L}{\sqrt{2\lambda}\sin(p/2)}}+O\left(\frac{1}{\sqrt{\lambda}}\right)\ .
\label{solmusub}
\ea
If we compare this result with the $\mu$-term for a giant magnon in ${\cal N}=4$ SYM \cite{GSV1}

\ba
\delta\epsilon^{\mu}_{SYM}&=&\left[-16\,g\sin^3\left(\frac{p}{2}\right)-\frac{16\sin\left(\frac{p}{2}\right)}{\pi}+8\,i\sin\left(\frac{p}{2}\right)-8\,i\sin(p)\right.\nonumber\\
&+&\left.\frac{4\,i\,L}{g}\cos\left(\frac{p}{2}\right)\right]e^{-2-\frac{L}{2g\sin(p/2)}}+O\left(\frac{1}{g}\right)\ ,
\label{musubSYM}
\ea
we notice that, differently to the classical contribution (Section 2), the substitution $g\rightarrow h(\lambda)$ is no longer enough to match the two results. In fact the relevant difference is the term proportional to $\sin(p)$, that is missing in (\ref{solmusub}) because of the different nature of the $S$-matrix contribution.
Furthermore it is very intersting to notice that the terms in (\ref{solmusub}) which are proportional to $c$ could be obtained by substituting $g/\sqrt{2}$ with $h(\lambda)$ the corresponding result of ${\cal N}=4$ SYM

\be
\delta\epsilon^{\mu}\simeq-\frac{16}{e^2}\,h(\lambda)\sin^3\left(\frac{p}{2}\right)e^{-\frac{L}{2h(\lambda)\sin(p/2)}}
\label{solmuSYM}
\ee
and then expanding $h(\lambda)$ as in (\ref{hsub}). Moreover, one can easily verify that this result can be obtained also by using the algebraic curve method as in \cite{GSV1}. In fact, in that paper the next-to-leading order of the finite-size correction to a ${\cal N}=4$ giant magnon was calculated, providing a beautiful matching between algebraic curve and L\"uscher approaches. In particular, following the first method, a parallel analysis leads to an expression for the one-loop energy shift that is given by the sum of various contributions

\be
\delta E=\delta E^{INT,(0)}+\delta E^{INT,(1)}+\delta E^{INT,(2)}+\delta E^{PL}+\delta E^{BP}+\delta E^{UP}\ .
\ee 
We will refer to \cite{GSV1} for the meaning of each factor in the previous equation.
Now, it is very reasonable that the sum of the last three terms $\delta E^{PL}+\delta E^{BP}+\delta E^{UP}$ cancels out and that $\delta E^{INT,(0)}, \delta E^{INT,(1)}$ give the same contribution as in SYM (see equations (5.10), (5.15), (5.18) and (5.21) in \cite{GSV1}). Hence we only need to express the term $\delta E^{INT,(2)}$ as

\be
\delta E^{INT,(2)}=\oint_{\mathbb{U}^+}\frac{dx}{2\pi i}\,\partial_x\Omega\left[\left(e^{i\frac{p}{2}}\,\frac{x-x_p^-}{x-x_p^+}+e^{i\frac{p}{2}}\,\frac{x-1/x_p^+}{x-1/x_p^-}\right)^2-4\right]e^{-\frac{i\,x\Delta}{h(\lambda)(x^2-1)}}
\label{FSU(2)xSU(2)}
\ee
in order to take into account the different structure of the ${\cal N}=6$ SCS $S$-matrix contribution, that is not given by the tensor product of the two $SU(2|2)$ invariant components as in SYM, but, as one can see for instance in (\ref{s-matrix2m}), by the multiplication of the single elements on the diagonal of $S^{AA}$ and $S^{BB}$.
The final result is

\ba
\delta E&\simeq&\oint_{\mathbb{U}^+}\frac{dx}{2\pi i}\left\lbrace \partial_x\Omega\left[\left(2\,\frac{x\,x_p^+-1}{x-x_p^+}\right)^2-4\right]e^{-ix\frac{\Delta}{h(\lambda)(x^2-1)}}\right\rbrace -\left[ \frac{16\sin\left(\frac{p}{2}\right)}{\pi}-8\,i\sin\left(\frac{p}{2}\right)\right. \nonumber\\
&+&\left. 16\,c\sin^3\left(\frac{p}{2}\right)-\frac{4\sqrt{2}L}{\sqrt{\lambda}}\left(i\cos\left(\frac{p}{2}\right)-2\,c\sin^2\left(\frac{p}{2}\right)\right)\right] e^{-2-\frac{L}{\sqrt{2\lambda}\sin(p/2)}}\ ,
\ea
where the first line coincides with the expression for the $F$-term (\ref{solF}) and the remaining terms are in perfect agreement with the result (\ref{solmusub}).
Actually, this matching at this moment is derived under the plausible arguments given above, which will deserve however more attention and a rigorous treatment in future work. 

For the $\mathbb{CP}^1$ giant magnon we follow the same steps of calculation again, hence we omit the details and give directly the result for the $\mu$-term up to $L/\lambda$ order:

\ba
\delta\epsilon^{\mu}_s&=&e^{-\frac{L}{\sqrt{2\lambda}\sin(p/2)}}\left[\frac{2\,i\sin(p/2)}{e}+\frac{1}{\sqrt{2\lambda}\,e\sin(p/2)}\left(\frac{2\,i}{\pi}-e^{i\frac{p}{2}}-1\right)+\right.\nonumber\\
&+&\left.\frac{L}{\lambda\,e}\left(2\,i\,c+\frac{\cos(p/2)}{\sin^2(p/2)}\right)\right]+O\left(\frac{1}{\lambda}\right)\ .
\ea
So, if we take the real part, for the considerations made in the previous Section, we have at this order the following not vanishing terms

\be
\mbox{Re}\left[\delta\epsilon^{\mu}_s\right]=e^{-\frac{L}{\sqrt{2\lambda}\sin(p/2)}}\left[-\frac{\cos(p/2)+1}{\sqrt{2\lambda}\,e\sin(p/2)}+\frac{L\cos(p/2)}{\lambda\,e\sin^2(p/2)}\right]+O\left(\frac{1}{\lambda}\right)\ .
\ee
Moreover, as far as the term proportional to $c$ is concerned, it could be simply obtained from the expansion of the leading term, if we suppose that its dependence on the coupling constant is given by the strong coupling expansion of $h(\lambda)$:

\be
\frac{2\,i\sin(p/2)}{e}\,e^{-\frac{L}{2h(\lambda)\sin(p/2)}}=\frac{2\,i\sin(p/2)}{e}\,e^{-\frac{L}{\sqrt{2\lambda}\sin(p/2)}}\left(1+\frac{c\,L}{\lambda\sin(p/2)}\right)+O\left(\frac{L}{\lambda^{3/2}}\right)\ .
\ee
Also in this case we can infer an heuristic derivation from the quantised algebraic curve approach \cite{GSV1} by suitably modifying the contribution, in the integral (\ref{FSU(2)xSU(2)}), which in the L\"uscher language is given by the $S$-matrix:

\ba
\delta E^{INT,(2)}&=&\oint_{\mathbb{U}^+}\frac{dx}{2\pi i}\,\partial_x\Omega\left[\left(e^{i\frac{p}{2}}\,\frac{x-x_p^-}{x-x_p^+}+e^{i\frac{p}{2}}\,\frac{x-1/x_p^+}{x-1/x_p^-}\right)-2\right]e^{-\frac{ix\Delta}{h(\lambda)(x^2-1)}}\\
&\simeq&\oint_{\mathbb{U}^+}\frac{dx}{\pi i}\left\lbrace \partial_x\Omega\left[\frac{x\,x_p^+-1}{x-x_p^+}-1\right]e^{-ix\frac{\Delta_s}{h(\lambda)(x^2-1)}}\right\rbrace+\frac{2\,i}{e}\sin\left(\frac{p}{2}\right)e^{-\frac{L}{\sqrt{2\lambda}\sin{p/2}}}\ .\nonumber
\ea 
We notice that for the single $SU(2)$ magnon we can correctly calculate via this formula only the leading $\mu$-term (\ref{musmall}), since it already is an $O(\lambda^0)$ contribution and then subleading in respect to the $F$-term. However, we have a nice confirmation of the result (\ref{musmall}).

Unfortunately, the comparison with the algebraic curve results initiated here seems not to give a check for the value of $c$, since it is enclosed in the definition of the $x^{\pm}_{p,q}$ variables, which are the constituent components of both - L\"uscher and algebraic curve - methods. Therefore we think that in these approaches $h(\lambda)$ behaves like a "mute" coupling constant, which can say nothing about its functional form, and then only the comparison with a string computation could decide something more.

\section{Conclusions}

In this paper we compute the classical and the first quantum finite-size corrections to the energy of giant magnons in the $SU(2)\times SU(2)$ sector of $\mathcal{N}=6$ superconformal Chern-Simons theory. Therefore we provide a check of the string result \cite{GHOS}, of the algebraic curve result \cite{IS}
and then a test for the all-loop Bethe Ansatz \cite{GV}, for the $S$-matrix \cite{AN} and, more generally, for the $AdS_4/CFT_3$ correspondence \cite{ABJM}.

We have proposed some generalised L\"{u}scher formulae heuristically derived from \cite{JL, GSV, BJ, HS}. We applied them in the one-particle case in order to calculate finite size correction to the energy of the so-called "small" giant magnon. It turns out a perfect agreement with algebraic curve calculations \cite{IS} for the $F$-term, while we propose a prediction for the $\mu$-term that needs some deeper understanding, as we explain in the main text.
For the giant magnon that lives on $\mathbb{R}\times S^2\times S^2$, we applied the formulae for the case of multi-particle states considering the strong coupling limit, where the interactions between elementary magnons is dominant and one can neglect the contributions coming from all the bound states of the theory. In particular, our $\mu$-term is in perfect agreement with the string result by \cite{GHOS} and the $F$-term matches the algebraic curve result by \cite{IS} for the "big" giant magnon. 

Indeed it would be extremely interesting to investigate for example the bound states $S$-matrix and the mirror \cite{AF} counterpart of the sector we considered, in order to study wrapping effects also at weak coupling (see \cite{BJ} for ${\cal N}=4$ SYM) and finite-size effects for dyonic giant magnons in $\mathbb{CP}^3$ (see \cite{ABR,R} for string computations and \cite{HS1} for string and L\"{u}scher-terms results in ${\cal N}=4$ SYM).

On the other hand, further investigations about sub-leading finite-size corrections at strong coupling could be - as mentioned above - an interesting future research direction as well.

\section*{Acknowledgements}
We are especially indebted to G. Grignani and R. Suzuki and we also thank F. Ravanini, M. Rossi, I. Shenderovich, R. Tateo for insightful discussions and correspondence.
We acknowledge the INFN grant "Iniziativa
specifica PI14" {\it Topics in non-perturbative gauge dynamics in
field and string theory} for travel financial support, and the University PRIN 2007JHLPEZ "Fisica Statistica dei Sistemi Fortemente Correlati all'Equilibrio e
Fuori Equilibrio: Risultati Esatti e Metodi di Teoria dei Campi". We thank the Galileo Galilei Institute for Theoretical Physics for the hospitality and
the INFN for partial support during the completion of this work.

\appendix
\section{Reconsidering the algebraic curve $F$-term for the "big" and "small" GM}

The equation (34) in \cite{IS} reads
\be
\delta\epsilon_{1-loop}=-\sum_{ij}\gamma_{ij}(-1)^{F_{ij}}\oint_{\mathbb{U}}\frac{dx}{4i}\frac{p'_i-p'_j}{2\pi}\cot\left(\frac{p_i-p_j}{2}\right)\Omega(x)
\label{int}
\ee
Let us consider this expression when $L$, namely $\Delta$, is large; then, since the quasimomenta have the following expression in terms of $\Delta$:
\be
p_i\simeq\frac{\Delta\,x}{x^2-1}\ ,
\ee
also the quasimomenta are large in this limit. Therefore we take the expansion of the cotangent when the $p_i$ are large and we ought to distinguish the two cases$x\in \mathbb{U}^{\pm}$:
\be
\cot\left(\frac{p_i-p_j}{2}\right)=\pm i\,(1+2\,e^{\mp i(p_i-p_j)}+...)\ ,
\label{exp}
\ee
where all the equilevel simbols are considered in the same expression.
Only the exponential part of (\ref{exp}) contributes in the integral (\ref{int}), then, after an integration by parts it becomes
\be
\delta\epsilon_{1-loop}=-\oint_{\mathbb{U}^+}\frac{dx}{4i\pi}\,\partial_x\Omega(x)\sum_{ij}\gamma_{ij}(-1)^{F_{ij}}e^{-i(p_i-p_j)}-\oint_{\mathbb{U}^-}\frac{dx}{4i\pi}\,\partial_x\Omega(x)\sum_{ij}\gamma_{ij}(-1)^{F_{ij}}e^{i(p_i-p_j)}\ ,
\ee
Now, one can easily verify, once made explicit the quasimomenta in terms of $x$ and $x^{\pm}$, that the two integrals above give the same contribution, in such way that one can perform the saddle-point evaluation on the same integral we obtain from the L\"{u}scher term calculations in the main text:
\be
\delta\epsilon_{1-loop}=-\oint_{\mathbb{U}^+}\frac{dx}{2i\pi}\,\partial_x\Omega(x)\sum_{ij}\gamma_{ij}(-1)^{F_{ij}}e^{-i(p_i-p_j)}
\ee
where
\be
\sum_{ij}\gamma_{ij}(-1)^{F_{ij}}e^{-i(p_i-p_j)}=4\,e^{-i\sqrt{2}\frac{(\Delta-L)}{\sqrt{\lambda}}}\left[\left(\frac{x-x^-_p}{x\,x^-_p-1}\right)^2-1\right]
\ee
in the case of the "big" GM, and
\be
\sum_{ij}\gamma_{ij}(-1)^{F_{ij}}e^{-i(p_i-p_j)}=4\,e^{-i\sqrt{2}\frac{(\Delta_{s}-L)}{\sqrt{\lambda}}}\left(\frac{x-x^-_p}{x\,x^-_p-1}-1\right)
\ee
in the case of the "small" GM. These expressions match exactly our $S$-matrix contributions in (\ref{intF}) and (\ref{sol1}).

\end{document}